**Increased GM-WM in a prefrontal network and decreased GM in the insula and the precuneus are associated with reappraisal usage: A data fusion approach**

Alessandro Grecucci[1, 2], Parisa Ahmadi Ghomroudi[1], Carmen Morawetz[3], Valerie Lesk[4], Irene Messina[5]

[1] Clinical and Affective Neuroscience Lab, Department of Psychology and Cognitive Sciences – DiPSCo, University of Trento, Rovereto, Italy

[2] Center for Medical Sciences – CISMed, University of Trento, Trento, Italy

[3] Department of Psychology, University of Innsbruck, Innsbruck, Austria

[4] Department of Psychology, University of Bradford, UK

[5] Universitas Mercatorum, Rome, Italy

**Corresponding author:**

Parisa Ahmadi Ghomroudi

Department of Psychology and Cognitive Sciences,

University of Trento,

Corso Bettini, 84, 38068,

Rovereto, Italy

**E-mail**: p.ahmadighomroudi@unitn.it

**Tel.** +39 04 64 808302

**Abstract**

Emotion regulation plays a crucial role in mental health, and difficulties in regulating emotions can contribute to psychological disorders. While reappraisal and suppression are well-studied strategies, the combined contributions of gray matter (GM) and white matter (WM) to these strategies remain unclear due to methodological limitations in previous studies. To address this, we applied a data fusion approach using Parallel Independent Component Analysis (Parallel ICA) to GM and WM MRI images from 165 individuals. Parallel ICA identified two networks associated with reappraisal usage. Network 1 included a large lateral and medial prefrontal cortical network, overlapping with the default mode network (DMN) and adjacent WM regions. Higher reappraisal frequency was associated with greater GM-WM density within this network, and this network was negatively correlated with perceived stress. Network 2 included the insula, precuneus, sub-gyral, and lingual gyri in its GM portion, showing a negative association with reappraisal usage. The WM portion, adjacent to regions of the central executive network (CEN), was positively associated with reappraisal usage. Regarding suppression, no significant network was associated with this strategy. This study provides new insights into individual differences in reappraisal use, showing a positive association between reappraisal frequency and increased gray and white matter concentration in a large frontal network, including regions of the frontal DMN and the CEN. Conversely, subcortical areas exhibited reduced gray and white concentration.

**Introduction**

Emotion regulation encompasses the processes through which individuals themselves can influence which emotions they experience, the timing of these emotions, and how they are felt and expressed (Gross, 1998b). Emotion regulation refers to the processes by which we control which emotions we have, when we have them, and how we experience and express them (Gross, 1998b). Deficits in emotion regulation or emotion dysregulation can result in disorders such as depression (LeMoult & Gotlib 2018), bipolar disorder (De Prisco et al., 2023, Lapomardi et al., 2021 a,b) substance-use disorder (Stellern et al., 2023), borderline personality disorder (Mendez-Miller et al.,.2022), eating (Walenda et al., 2021; Munguia et al., 2021) and gambling disorders (Munguia et al., 2021) insomnia (Palagini et al., 2017) and somatic symptom disorder (Schnabel, 2022). Two well-studied emotion regulation strategies are *reappraisal* and *suppression* (Gross, 1998). *Reappraisal* is an antecedent-focused strategy that aims to modify emotional responses before they are fully activated. This approach involves reinterpreting the meaning of a situation in order to alter the emotional impact it has (Gross, 1998). In contrast, *suppression* is a response-focused strategy and is defined as the attempt to inhibit ongoing emotional expressions (Gross & Levenson, 1993). Empirical studies exploring reappraisal and suppression indicate that reappraisal effectively reduces negative emotional expressions without increasing physiological arousal, even compared to instances where no regulation is applied (Gross, 2002, 2015; Goldin et al., 2019). By contrast, suppression tends to reduce the experience of positive emotions without significantly altering negative emotions, and it has been associated with increased physiological arousal, including heightened cardiovascular activity, increased sympathetic nervous system activation, and elevated stress responses (Gross & Levenson, 1993, 1997; Gross, 2002; Mauss et al., 2005; Brans et al., 2013).

Over the past two decades, researchers have sought to understand the neural bases of specific emotion regulation strategies. It is believed that distinct brain regions are responsible for the different strategies involved in emotional regulation. For instance, the dorsolateral PFC (dlPFC), medial PFC (mPFC), ventrolateral PFC (vlPFC), and dorsal anterior cingulate cortex (ACC) are thought to be recruited during reappraisal (Nelson et al., 2015), with this process also leading to a reduction in activity within limbic areas (Buhle et al., 2013). A meta-analysis of 48 task-related fMRI studies on reappraisal (Buhle et al., 2014) confirmed this model by showing increased activation in the bilateral dorsolateral and ventrolateral prefrontal cortex (dlPFC, vlPFC), dorsal anterior cingulate cortex (dACC), supplemental motor area (SMA), and inferior/superior parietal cortex during both upregulation and downregulation of emotion. Another meta-analysis revealed that reappraisal enhanced connectivity between the right dorsolateral PFC, left ventrolateral PFC, dorsomedial PFC, and subcortical regions (Berboth & Morawetz, 2021). A more recent meta-analysis of 42 fMRI studies on reappraisal and acceptance showed decreased activation in limbic areas and increased activity in the dlPFC and left lateral PFC during reappraisal (Monachesi et al., 2023). Using TMS-fMRI, He et al. (2023) demonstrated that facilitation of the ventrolateral PFC led to greater activation of both the ventrolateral and ventromedial PFC during reappraisal compared to no regulation. Moreover, they found that prefrontal-subcortical couplings through the ventromedial PFC were stronger during reappraisal, accompanied by reduced activity in the amygdala and insula. Interestingly, large portions of the prefrontal cortex, particularly the medial regions, overlap with the default mode network (DMN). The DMN includes the medial PFC, anterior and posterior cingulate cortex (ACC/PCC), precuneus (PCC), bilateral inferior parietal cortices (Garrity et al., 2007; Raichle et al., 2001; Supekar et al., 2010; Whitfield-Gabrieli & Ford, 2012), and the medial temporal lobe (Buckner et al., 2008). The DMN is thought to be active during periods of rest (Buckner, 2013). Regarding emotion regulation, certain DMN regions are believed to contribute to reappraisal. For example, reduced resting-state functional connectivity between the right amygdala, medial PFC, and posterior cingulate cortex was found to predict success in using reappraisal strategies (Uchida et al., 2015). Additionally, two resting-state functional connectivity studies reported that individual differences in reappraisal influence the DMN's functional connectivity (Martins & Mather, 2016; Morawetz et al., 2016). Collectively, these findings suggest that reappraisal relies on a prefrontal network that likely overlaps with the frontal hub of the DMN. Furthermore, portions of the Central Executive Network (CEN), particularly in the dorsal frontal regions, also appear to be involved.

      Besides the functional aspects of reappraisal, another line of research has tried to understand its structural basis. One sMRI study using region of interest (ROI) analysis found a positive correlation between the volume of the dorsal anterior cingulate cortex and the use of reappraisal (Giuliani et al., 2011a). Similarly, a voxel-based morphometry (VBM) study reported a positive correlation between reappraisal and the volume of the right and left amygdala (Hermann, Bieber et al., 2013a). In another study, Pappaianni and colleagues (2020) applied an unsupervised machine learning technique called source-based morphometry (SBM; Xu et al., 2009) to investigate individual differences in structural brain features associated with reappraisal usage. Their findings revealed gray matter differences in a network consisting of frontal, temporal, and parietal regions between low and high reappraisers. In a subsequent study, increased gray matter concentration in a temporo-parahippocampal-orbitofrontal network, partially overlapping with the DMN, was shown to successfully predict individual differences in the use of reappraisal (Ahmadi Ghoumroudi et al., 2023).

Regarding suppression, fewer studies have investigated this. One functional imaging study found a positive correlation between the right dorsomedial prefrontal cortex (dmPFC) volume and use of suppression (Kühn et al., 2011). Another study observed a positive correlation between anterior insula volume and suppression use (Giuliani et al., 2011b). Additionally, increased activity in the PFC, insula, and amygdala was reported during suppression, along with a positive association between the amygdala and dorsal ACC, and a negative association between the left centromedial amygdala and the supplementary motor area (Pico-Perez et al., 2018). Finally, from a structural point of view, a study by Ahmadi Ghomroudi and colleagues (2023) identified networks with higher gray matter concentration within the insular network and fronto-parietal-cerebellar network to be predictive of suppression.

Although the structural studies on reappraisal and suppression generally align with functional findings, it is important to acknowledge that each provides only partial confirmation on the neural bases of these strategies, likely due to inherent methodological limitations. The majority of these studies have used univariate approaches (such as VBM) or have a priori selected regions of interest (ROI) analyses that do not show the whole picture regarding the neural bases of these strategies.

In recent years, affective neuroscience has increasingly used machine learning techniques due to their enhanced ability to capture complex interactions across multiple brain regions compared to traditional statistical approaches. These methods can identify subtle, widespread changes in brain structure and function that may go undetected with conventional analyses (Ahmadi Ghomroudi et al., 2023; Ahmadi Ghomroudi et al., 2024; Grecucci et al., 2022; Hebart & Baker, 2018). Specifically, machine learning examines the relationships among groups of voxels simultaneously, rather than analyzing each voxel independently, thus offering a more integrative view of brain activity (Hebart & Baker, 2018; Grecucci et al., 2022). Recent work has provided fresh insight into the functional and neural bases of emotion regulation strategies, revealing distinct brain networks associated with reappraisal, suppression, and acceptance. Specifically, reappraisal has been linked to a temporo-parahippocampal-orbitofrontal network, suppression to insular and fronto-temporo-cerebellar networks, and acceptance to decreased affective and executive network activity alongside increased sensorimotor activity (Ahmadi Ahmadi Ghomroudi et al., 2023; Ahmadi et al., 2024). However, a key limitation of current machine learning studies in emotion regulation research is their focus on a single neuroimaging modality at a time, typically resting-state functional connectivity or gray matter volume. While these studies have provided valuable insights into functional and structural correlates of emotion regulation, they fail to capture how these components interact. Furthermore, few studies have examined the joint contributions of gray and white matter (Baggio et al., 2023; Grecucci et al., 2025), despite evidence suggesting that structural connectivity plays a crucial role in shaping individual differences in emotion regulation. Moreover, many studies rely on predefined regions of interest or task-based paradigms, which may overlook more subtle and distributed neural mechanisms underlying emotion regulation strategies.

To address these gaps, we used unsupervised machine learning technique, a powerful data-driven approach that can uncover hidden patterns in large, complex datasets without the need for predefined labels (Vieira, Pinaya, & Mechelli, 2020). Unlike traditional analyses, unsupervised learning enables the identification of naturally occurring brain networks, offering a biologically plausible, integrative perspective on neural mechanisms. Additionally, this approach allows for data fusion techniques (Baggio et al., 2023; Grecucci et al., 2025), which integrate information from multiple sources, such as gray matter and white matter, providing a more comprehensive view of brain structure and function.

**The present study.**

This study aimed to study the neural substrates of reappraisal and suppression through data fusion unsupervised machine learning. By applying unsupervised learning, we aimed to segment the brain into independent networks of covarying gray matter and white matter features and subsequently predict which of these networks were associated with reappraisal and suppression strategies. This dimensionality reduction approach has the benefits of limiting computational demands, minimizing redundant data, focusing on relevant information, and reducing the risk of overfitting (Guyon & Elisseeff, 2003). To achieve this, we used parallel independent component analysis (pICA) (Liu et al., 2007; Yang et al., 2019; Baggio et al., 2023) to decompose brain data into distinct neural networks. This approach compresses the high-dimensional voxel space into a manageable number of biologically plausible networks, which are derived from structural and functional covariation rather than static atlas-based regions (Ahmadi Ghomroudi et al., 2023; Grecucci et al., 2022). This approach aligns with the understanding that emotion regulation strategies are supported by distributed networks rather than isolated brain regions (Biswal et al., 2010; Morawetz et al., 2017; Ahmadi Ghomroudi et al., 2023; Grecucci et al., 2024)

In our analyses, we included both gray matter and white matter components, given that both types of brain tissue are influenced by shared environmental and genetic factors (Baggio et al., 2023; Spalletta, Piras, & Gili, 2018) and are therefore important to investigate. Of note, WM contribution to reappraisal and suppression has never been explored as far as we are aware. From a methodological point of view, WM integrity is often measured using diffusion tensor imaging (DTI), which assesses the microstructural integrity of WM fibers by tracking water molecule displacement (Assaf & Pasternak, 2008; Alba-Ferrara & de Erausquin, 2013). However, DTI's high sensitivity to noise can lead to issues with the reproducibility in regions of interest (Radwan et al., 2022). Incorporating pure WM concentrations in conjunction with GM data enhances our ability to detect distributed WM alterations across networks, without the constraints of specific fiber tracts (Baggio et al., 2023).

      We hypothesize that reappraisal usage can be predicted by a PFC network that might also overlap with the DMN. Medial frontal regions may facilitate reappraisal by enhancing introspective and self-referential processing (Grecucci, 2013a; Ahmadi Ghomroudi et al., 2023). We predict a positive relationship between reappraisal usage and increased GM-WM in frontal regions. Because of the nature of reappraisal, and the involvement of top-down mechanisms, we also predict that brain regions that are partially connected with executive control regions, specifically within the dorsal and ventral parts of the PFC, which are essential for control (Ahmadi Ghomroudi, 2023; Grecucci, 2013a) will be involved. In addition, based on previous studies (Ahmadi Ghomroudi, 2023) we hypothesize that suppression usage will be associated with increased gray and white matter concentrations within the somatosensory network, particularly in the insula. The insula, which is central to bodily and emotional awareness, is thought to be a key initiator of suppression processes. In addition, based on the previous findings (Niederhauser, Sefidan, & Annen, 2021) that showed reappraisal is negatively associated with perceived stress, our study aims to further investigate whether the neural circuits involved in reappraisal usage are directly linked to stress. We predict this to be the case.

**Method**

**Participants**

One hundred sixty-two participants (55 females; mean age = 37.47 years, SD = 19.74) participated in this study. Emotion regulation was assessed using the emotion regulation questionnaire (ERQ), with a mean reappraisal score of 4.60 ± 0.95 and a mean suppression score of 3.76 ± 1.11. The data were obtained from the 'Leipzig Study for Mind-Body-Emotion Interactions' (OpenNeuro database, accession number ds000221, LEMON), collected at the Max Planck Institute for Human Cognitive and Brain Sciences (MPI CBS) in Leipzig (Babayan et al., 2019). Data collection adhered to the Declaration of Helsinki, and the study protocol was approved by the Ethics Committee of the University of Leipzig Medical Faculty (reference number 154/13-ff). Inclusion criteria included the absence of cardiovascular, psychiatric, or neurological disorders, and no malignant diseases or certain medications. Individuals with reported drug or excessive alcohol use were excluded. All participants provided written informed consent for anonymous data sharing and were compensated upon completion of assessments

**Behavioural data**

The German version of the ERQ (Abler and Kessler, 2009), was used to measure the frequency of usage of two reappraisal and suppression. This questionnaire consists of ten items, with six items assessing reappraisal and four items assessing suppression. Each item is rated on a 7-point Likert scale, ranging from 1 (strongly disagree) to 7 (strongly agree). Additionally, the Perceived stress questionnaire (PSQ, Levenstein et al., 2003) was used to assess the stress level of participants. The Perceived Stress Questionnaire (PSQ; Levenstein et al., 1993) is a 30-item tool used to assess stress levels, particularly in chronic pain patients. Participants respond using a 4-point scale: 1 = Almost Never, 2 = Sometimes, 3 = Often, and 4 = Usually. The PSQ included these subscales: Demands, Lack of Joy, Worries, and Tension.

**Image acquisition**

Structural MRI scans were acquired using a 3 Tesla MAGNETOM Verio scanner (Siemens Healthcare GmbH, Erlangen, Germany) equipped with a 32-channel head coil. During the data acquisition period, no significant maintenance or updates were performed on the scanner that could have affected data quality. Structural volumes were captured in 176 slices over a total scanning time of 8 minutes and 22 seconds.

**Preprocessing**

First, structural MRI data quality was checked to identify and exclude any potential artifacts. Preprocessing was then performed using the Computational Anatomy Toolbox (CAT12; http://www.neuro.uni-jena.de/cat/) within the Statistical Parametric Mapping (SPM12) framework in MATLAB (The Mathworks, Natick, MA). Structural images were manually aligned with the anterior commissure set as the origin. Next, CAT12 segmented the images into gray matter, white matter, and cerebrospinal fluid. Diffeomorphic Anatomical Registration using Exponential Lie algebra (DARTEL) tools for SPM12 were used for image registration, replacing traditional whole-brain registration methods (Yassa and Stark, 2009; Grecucci et al., 2016; Pappaianni et al., 2018). Finally, the DARTEL-registered images were normalized to MNI-152 standard space, and each image was smoothed with a 12-mm full-width at half-maximum (FWHM) Gaussian kernel [12, 12, 12].

**Data fusion unsupervised machine learning approach**

Paralle ICA was applied to the structural MR images to identify independent circuits across the whole brain (Xu et al., 2009; Pappaianni et al., 2018; 2020; Sorella et al., 2019; Saviola et al., 2020; Baggio et al., 2023). In this study, the Fusion ICA Toolbox (FIT, available at http://mialab.mrn.org/software/fit) inside MATLAB environment (The Mathworks, Natick, MA). The number of independent components (ICs) for the analysis was determined using information-theoretic criteria to estimate the optimal model complexity (Wax & Kailath, 1985). This approach ensures that the balance between data fit and model simplicity is maintained, avoiding overfitting. To further validate the stability and consistency of the extracted components, ICASSO was run ten times, using the Infomax algorithm for component extraction. This procedure helped confirm the reliability of the selected number of ICs for subsequent analysis (Himberg et al., 2003; Himberg, Hyvärinen, & Esposito, 2004). Then, Stepwise regression analysis was conducted using the regression module in JASP (Version 0.16.0, JASP Team, 2021; Fig. 1). Reappraisal and suppression scores were entered as dependent variables in two separate regressions, with the loading coefficients of the ICs as independent variables.

**Results**

The Minimum Description Length (MDL) algorithm estimated the presence of 17 GM and 11 WM networks in the data. These networks were estimated accordingly via pICA. Multiple linear regression using stepwise regression analysis revealed that, GM1, GM12, GM15, GM17, WM2, WM3, WM5, and WM11 significantly predicted reappraisal usage, $F(10, 151) = 3.396$, $p < 0.001$. See table 1. However, only two pairs of GM and WM networks were significantly correlated: Network 1 included GM17 and WM11. GM17 was negatively correlated with WM11 (r=-0.922, p<0.001). Network 2 including GM12 and WM5. GM12 positively correlated with WM5 (r=0.580, p<0.001). Since we were interested in finding joint GM-WM networks to predict reappraisal, we discarded the spurious networks (ICs with just GM or WM significant) and focused only on the pairs in which both the GM and the WM were significant. Of note GM17 was also negatively with WM5 (r=-0.178, p=0.24). See Table 2 for the correlations between modalities.

-------------------------------------------------------

Please insert Tables 1 and 2 about here

-------------------------------------------------------

Network 1 consisted of GM17 and WM11. GM17 included large portions of medial and lateral prefrontal areas, the anterior cingulate, the fusiform are and portions of the occipital cortex. Of GM17 was positively associated with reappraisal indicating increased GM for higher reappraisal usage. Of note GM17 negatively correlated with the PSQ questionnaire subscales: Worries (r=-0.161, p=0.041), Tension (r= -0.179, p=0.023) and Demands (r= - 0.232, p= 0.003). WM11 was positively associated with reappraisal and included large white matter portions adjacent to the thalamus, the caudate, temporal poles, and fronto-parietal regions possibly associated with the Central-Executive network. Network 2 consisted of GM12 and WM5. GM12 included the posterior part of the insula, the precuneus, the lingual gyrus and the subgyral. GM12 was negatively associated with reappraisal, meaning that decreased GM was found for high reappraisal usage. WM5 included orbito-frontal and occipital regions and was positively associated with reappraisal. In sum, GM17, WM11 and WM5 including regions of both the DMN and the ECN were positively associated with reappraisal, and GM12, mainly subcortical negatively associated. See table 3 and 4 for the list of brain regions involved in Network 1 and Network 2 and Figure 1 and 2 for a graphical representation of results. Regarding suppression, multiple linear regression using stepwise regression analysis returned a non-significant model, $F(10, 160) = 2.923$, $p = 0.089$.

-------------------------------------------------------

Please insert Tables 3 and 4 about here

-------------------------------------------------------

**Discussion**

The aim of this study was to provide new evidence regarding how individual differences in the use of two distinct emotion regulation strategies, namely reappraisal and suppression can be predicted by stable structural features of the brain and their relationship with perceived stress. To achieve this, we applied a data fusion unsupervised machine learning algorithm, known as parallel ICA, to sMRI scans from 162 healthy individuals. This method decomposed the brain into independent networks of covarying gray and white matter, returning 17 GM and 11 WM networks. A stepwise multiple linear regression analysis was then conducted to identify which networks are predictive of these strategies. Among the networks, only two GM-WM network pairs showed significant correlations with reappraisal. Unfortunately, we could not find any significant result for suppression.

The first network (Network 1) included GM17 and WM11. GM17 consisted of extensive medial and lateral prefrontal areas, the anterior cingulate, the fusiform area, and portions of the occipital cortex. GM17 was predictive of reappraisal, such that the higher the GM density in these regions the greater use of reappraisal. These results align and confirm previous functional studies on reappraisal. The actual model of the brain bases of reappraisal indeed include areas such as the dorsolateral PFC (dlPFC), the medial PFC (mPFC), the ventrolateral PFC (vlPFC), and the dorsal anterior cingulate cortex (ACC) (Nelson et al., 2015; Buhle et al., 2013; Morawetz et al.,2017; Morawetz et al., 2020). This was confirmed by three recent meta-analyses on this topic. A meta-analysis by Buhle et al (2014), showed that reappraisal is associated with increased activation in the bilateral dlPFC, vlPFC, dorsal ACC (dACC), supplemental motor area (SMA), and parietal regions during both emotional upregulation and downregulation (Buhle et al., 2014). Similarly, enhanced connectivity between prefrontal and subcortical regions, including the right dlPFC, left vlPFC, and dorsomedial PFC, was observed during reappraisal (Berboth & Morawetz, 2021). Another more recent meta-analyses further highlight increased prefrontal engagement, particularly in the dlPFC and lateral PFC (Monachesi et al., 2023).

Notably, the more medial part of the PFC that was included in GM17, greatly overlaps with the default mode network (DMN), which includes the anterior/posterior cingulate cortex (ACC/PCC), precuneus, inferior parietal cortices, and medial temporal lobe (Buckner et al., 2008). DMN regions appear to support reappraisal, as reduced connectivity between the amygdala, medial PFC, and PCC predicts reappraisal success (Uchida et al., 2015). Moreover, individual differences in reappraisal usage influence DMN connectivity (Martins & Mather, 2016; Morawetz et al., 2016).

Moreover, the more lateral part of the PFC included in GM17, extends to regions usually ascribed to the frontal part of the CEN. This was also confirmed by the white matter portions of the WM11, which positively predicted reappraisal too. WM11 included large white matter areas not only adjacent to the thalamus, and caudate, but also to fronto-parietal regions, mainly belonging to the Central Executive Network (CEN).

Collectively, these findings align with the notion that reappraisal is a multi-component process involving both top-down cognitive control and bottom-up affective processing (Ochsner & Gross, 2008; Feldman Barrett & Satpute, 2013). Specifically, the frontal hubs identified in GM17 and WM11 likely support distinct sub-processes of reappraisal—such as evaluating emotional meaning (DMN regions) and actively transforming those appraisals (CEN regions). The habitual use of reappraisal may, over time, shape the structural integrity of these interconnected networks, reflecting greater efficiency in managing emotional experiences. Future studies should examine how reappraisal success (i.e., objective measures of effective emotion regulation) relates to the structural and functional properties of these networks, offering a more precise link between habitual reappraisal and real-time regulatory outcomes (Uchida et al., 2015; Martins & Mather, 2016; Morawetz et al., 2016).

Moreover, GM17 as expected was negatively correlated with perceived stress (PSQ). The higher the GM17 grey matter concentrations, the more frequent the usage of reappraisal and the less the perceived stress, confirming a appositive role of reappraisal in reducing stress experience. Previous research has already established a link between reappraisal usage and perceived stress (Niederhauser, Sefidan, & Annen, 2021). Our study confirms and expands this by connecting reappraisal and stress with the positive role of a prefrontal network in producing such effect.

The second network (Network 2) consisted of GM12 and WM5. GM12 included the posterior insula, precuneus, lingual gyrus, and subgyral regions. In contrast to GM17, GM12 was negatively predictive of reappraisal, indicating that the lower the GM density in these regions the higher the use of reappraisal.

The posterior insula has related to emotions and emotion regulation in multiple ways. For example, decreased activity in the insula and in the amygdala during reappraisal in response to unpleasant stimuli is a common finding (Goldin et al., 2008). Previous studies on emotion regulation in interpersonal contexts reported significant modulation of insular activity during the regulation process (Grecucci et al., 2013a, b). The authors proposed that this modulation of insular activity may reflect the regulation of emotion-induced physiological arousal. Of note, the insula is a major hub of the salience network (Smith and Lane, 2015; Morawetz et al., 2020). Muhtadie et al. (2021) proposed that the salience network, particularly the insular cortex, plays a central role in processing emotional salience. The insula integrates input from various brain regions, such as the amygdala, anterior cingulate cortex, and hypothalamus, to create a representation of the body's affective and proprioceptive states. The insula plays a pivotal role in processing sensory information from both internal and external environments to create a unified and conscious representation of one's emotional state (e.g., Zaki et al., 2012). It is also critical for mapping the arousal associated with emotional experiences (Grecucci et al., 2013a, b). Alongside the insula, the ventrolateral prefrontal cortex (VLPFC) has been consistently linked to successful emotion regulation across various strategies (Morawetz et al., 2017; Li et al., 2021), including acceptance (Messina et al., 2021).

Based on this, we suggest that the effect of reappraisal usage may reduce activity and subsequently the morphometry of the insula. This result aligns with studies on other emotion regulation strategies (e.g. acceptance), in which a consistent pattern of decreased brain activity in regions associated with emotional processing such as the precuneus, and the insula (Messina et al. 2023).

The precuneus was indeed included in GM12. The precuneus plays a role in various functions, such as autobiographical memory, self-processing, and consciousness, while also supporting higher-order cognition as a posterior hub of the Default Mode Network (Cavanna et al., 2006; Tanglay et al., 2022). Anatomical and functional studies have indicated the involvement of the precuneus in anxiety disorders, including panic disorder, social anxiety disorder, and generalized anxiety disorder (O'Neill et al., 2015; Baggio et al., under review). It is not surprising therefore that we found a negative relationship with reappraisal usage, as previous studies clearly reported an inverse relationship between reappraisal and anxiety (Ahmadi Ghoumroudi et al., 2023).

Concerning the occipital areas included in GM12, previous studies on the perception of emotional stimuli have reported enhanced activation inside the lingual gyrus that becomes reduced when individuals apply reappraisal ((Li et al., 2018; Ochsner et al., 2002; Goldin et al., 2008; Sun et al., 2022). Other studies have confirmed this engagement of occipital regions during reappraisal (Ochsner et al., 2004b; Urbain et al., 2017). The presence of occipital regions in the reappraisal network suggests that reappraisal is not solely a language-based process but also critically relies on visual imagery. Regions such as the lingual gyrus, inferior occipital gyrus, and cuneus are typically active during vivid visual processing (Daselaar et al., 2010; Zvyagintsev et al., 2013) however, our structural findings indicate reduced gray matter density in these areas. In parallel, the increased white matter concentration in WM5 observed adjacent to these occipital areas suggests enhanced connectivity between visual cortices and frontal control regions. The involvement of visual regions in reappraisal may be related to imagery techniques in psychotherapy for which the client is guided through a modification of traumatic experiences not only using language-based techniques, but also imagery. One case in point the so called imagery rescripting, a therapeutic technique in which patients reimagine and transform distressing memories to reduce their emotional impact (Young, 1990; Young et al., 2003; Arntz, 2012;). Recent investigations into the neural correlates of visual imagery vividness further support the involvement of these occipital regions in generating detailed mental images (Kvamme et al., 2024; Fulford et al., 2017). Together, these findings underscore that reappraisal, visual imagery, and imagery rescripting depend on integrated visual imagery processes, highlighting the critical role of occipital regions not only in generating vivid mental images but also in supporting effective emotion regulation

The role of occipital regions is also confirmed by the presence of large white matter portions adjacent to the occipital lobes included in the WM5 part of Network 2. WM5 also includes portions of WM adjacent to the orbifrontal cortex. Previous studies have highlighted the role of the orbitofrontal network in successfully predicting the individual differences in the use of reappraisal (Ahmadi Ghoumroudi et al., 2023; 2024). Higher grey matter concentration within a temporo-parahippocampal-orbitofrontal network was found to predict reappraisal usage (Ahmadi Ghoumroudi et al., 2023; 2024). Of note, we found a positive relationship between WM5 and reappraisal usage. The orbitofrontal cortex is a key frontal region implicated in successful reappraisal (Wager et al., 2008). Research suggests that the OFC contributes to cognitive control functions essential for reappraisal, such as inhibition (Ochsner et al., 2004; Banks et al., 2007; Kanske et al., 2011). Notably, recent studies underscore the significance of connectivity between the OFC and the amygdala - a critical region for emotion processing—in the context of reappraisal. For instance, Gao et al. (2021) demonstrated that functional coupling between the OFC and amygdala is associated with the use of reappraisal. Similarly, Kanske et al. (2021) reported decreased amygdala activity and increased OFC activity during reappraisal. These findings suggest that the interplay between the OFC and amygdala may serve as a fundamental neural mechanism underpinning emotional regulation strategies, particularly reappraisal (Ahmadi Ghoumroudi et al., 2023; 2024).

The second aim of the present study was to detect covarying GM-WM networks to predict suppression usage. We could not find any network to be related to this strategy and relatively few studies have examined the nature of suppression as an emotion regulation strategy. Previous studies have reported less brain modification in suppression compared to reappraisal (Goldin et al., 2008). One functional study reported a positive correlation between the volume of the right dorsomedial prefrontal cortex (dmPFC) and suppression usage (Kühn et al., 2011). Another study found a similar positive correlation between anterior insula volume and the use of suppression (Giuliani et al., 2011b). During suppression, increased activity has been observed in the PFC, insula, and amygdala, accompanied by a positive association between the amygdala and dorsal anterior cingulate cortex (ACC), and a negative association between the left centromedial amygdala and the supplementary motor area (Pico-Perez et al., 2018). From a structural perspective, a study by Ahmadi Ghomroudi and colleagues (2023) identified that networks with greater gray matter concentration in the insular network and the fronto-parietal-cerebellar network are predictive of suppression use. These findings collectively highlight the involvement of both structural and functional neural mechanisms in suppression. Therefore, these findings do not reveal a clear picture of the neural basis of suppression. In fact, the strategy of suppression often yields less consistent neural findings compared to other emotion regulation strategies, such as reappraisal, for several reasons. One possibility is that suppression encompasses multiple subprocesses, including inhibiting expressive behavior, redirecting attention, and modulating physiological responses. These varied components may activate different and sometimes overlapping neural regions, making it difficult to identify a distinct neural signature. Another possibility is that suppression occurs after an emotional response has been generated, focusing on inhibiting the outward expression rather than altering the emotional experience itself. This downstream approach might rely more on the modulation of peripheral systems (e.g., motor and physiological responses) and less on the centralized cognitive and emotional control networks typically studied in neuroimaging. Furthermore, suppression often recruits compensatory mechanisms, such as increased activation in control-related regions (e.g., prefrontal cortex) to counteract heightened emotional responses in areas like the amygdala. Another possibility may reside in the fact that suppression may be more difficult to detect by self-reported questionnaires as the one used in the present study (ERQ).

**Conclusions and limitations**

This study aimed to investigate whether the use of reappraisal and suppression as emotion regulation strategies could be predicted from grey and white matter networks. We found two networks to be involved in reappraisal but none in suppression. GM17, WM11, and WM5 including regions associated with the Default Mode Network (DMN) and the Executive Control Network (ECN) were positively predictive of reappraisal, suggesting that increased structural features in these areas are linked to higher reappraisal usage. Conversely, GM12, primarily subcortical, was negatively predictive of reappraisal, indicating that reduced structural features in these regions may facilitate this emotion regulation strategy. Furthermore, GM17 was negatively related to stress.

This study does not come without limitations. The use of self-report questionnaires may introduce biases, and the study's methodological approach, while innovative, leaves room for future research to explore the comparative effectiveness of different machine-learning algorithms. Further, incorporating a task-based measure of regulation success and supplementing self-report measures with additional questionnaires could provide a more comprehensive assessment of regulatory processes. Future studies should incorporate additional questionnaires specifically designed to assess visual imagery vividness and reappraisal tendencies. Such measures would allow researchers to directly correlate subjective reports with neural markers, thereby clarifying whether individual differences in visual imagery capability predict reappraisal usage and efficacy and ultimately guiding more targeted therapeutic interventions.

Additionally, although the sample size is consistent or even larger compared with recent structural neuroimaging studies (e.g., Picó-Pérez et al., 2019; Baltruschat et al., 2021) this can be expanded upon in the near future. Although we used a data fusion approach to detect GM-WM covariances for the first time, the role of other neuroimaging modalities needs to be explored, especially concerning merging structural with functional modalities. Future studies may want to explore this aspect. Finally, our analyses did not yield a significant model for suppression, indicating no clear relationship between suppression and the identified brain networks. Future studies may want to understand this better.

**Data Availability:**

The LEMON dataset is publicly available via the Gesellschaft für wissenschaftliche Datenverarbeitung mbH Göttingen (GWDG) (https://www.gwdg.de/). Both raw and preprocessed data can be accessed either through a web browser at https://ftp.gwdg.de/pub/misc/MPI-Leipzig_Mind-Brain-Body-LEMON/ or via FTP at ftp://ftp.gwdg.de/pub/misc/MPI-Leipzig_Mind-Brain-Body-LEMON/. Should the dataset be relocated in the future, it will remain accessible using the persistent identifier PID 21.11101/0000-0007-C379-5, which can be resolved at http://hdl.handle.net/21.11101/0000-0007-C379-5.

**Tables**

| Table 1. Multiple regression results | | | | | |
|---|---|---|---|---|---|
| **Variable** | **β** | **SE** | **t** | **p** | **95% CI** |
| picaGM1 | 8,387 | 3,117 | 2,691 | 0,008 | [2.249, 14.525] |
| picaGM4 | 6,666 | 3,425 | 1,946 | 0,054 | [−0.066, 13.398] |
| picaGM12 | -10,765 | 4,346 | -2,477 | 0,014 | [−19.342, −2.188] |
| picaGM14 | -4,479 | 2,514 | -1,782 | 0,077 | [−9.437, 0.479] |
| picaGM15 | 14,629 | 4,48 | 3,265 | 0,001 | [5.771, 23.487] |
| picaGM17 | 10,402 | 3,387 | 3,071 | 0,003 | [3.707, 17.097] |
| picaWM2 | 10,714 | 3,225 | 3,322 | 0,001 | [4.336, 17.092] |
| picaWM3 | -18,672 | 8,156 | -2,289 | 0,023 | [−34.793, −2.551] |
| picaWM5 | 12,06 | 4,49 | 2,686 | 0,008 | [3.196, 18.924] |
| picaWM11 | 11,787 | 5,582 | 2,112 | 0,036 | [0.795, 22.779] |

CI = Confidence Interval, SE = standard error

| Table 2. Correlations between modalities. | | |
|---|---|---|
| **GM** | **WM** | **Correlation** |
| 16 | 8 | 0,9387 |
| **17** | **11** | **-0,922** |
| 15 | 9 | -0,9209 |
| **12** | **5** | **0,5801** |
| 8 | 1 | -0,5223 |
| 13 | 7 | -0,4534 |
| 7 | 11 | 0,4156 |
| 14 | 2 | 0,3431 |
| 2 | 11 | -0,3237 |
| 6 | 11 | -0,3188 |
| 11 | 4 | 0,3052 |
| 1 | 4 | -0,2811 |
| 3 | 6 | -0,2755 |
| 10 | 11 | -0,2529 |
| 4 | 11 | -0,248 |
| 9 | 7 | 0,2277 |
| 5 | 2 | -0,2068 |

Networks GM17-WM11, and GM12-WM5 are correlated and significant predictors of Reappraisal.

**Table 3. Brain regions of Network 1**

| Network 1 | Area | Brodmann Area | Volume (cc) | Peak (x, y, z) |
|---|---|---|---|---|
| **GM 17** | | | | |
| | Anterior Cingulate | 10, 24, 25, 32 | 5.0/6.8 | 7.2 (-1, 35, 0) / 7.9 (3, 33, -5) |
| | Medial Frontal Gyrus | 9, 10, 11, 25, 32 | 1.1/5.0 | 4.4 (-1, 31, -15) / 6.4 (9, 38, -7) |
| | Extra-Nuclear | * | 0.3/0.5 | 5.1 (-1, 29, 0) / 6.2 (3, 32, 3) |
| | Lingual Gyrus | 17, 18 | 1.0/1.7 | 5.6 (-1, -85, -10) / 5.8 (1, -85, -10) |
| | Subcallosal Gyrus | 11, 25 | 0.0/0.6 | -999.0 (0, 0, 0) / 5.4 (6, 24, -11) |
| | Sub-Gyral | * | 1.2/3.2 | 4.0 (-28, -52, 34) / 4.9 (12, 39, 1) |
| | Inferior Frontal Gyrus | 10, 11, 47 | 0.0/0.7 | -999.0 (0, 0, 0) / 4.9 (15, 18, -17) |
| | Rectal Gyrus | 11 | 0.0/0.8 | -999.0 (0, 0, 0) / 4.8 (12, 19, -20) |
| | Middle Frontal Gyrus | 10 | 0.3/3.0 | 3.3 (-33, 20, 28) / 4.2 (25, 51, 3) |
| | Fusiform Gyrus | 18, 19, 20 | 0.2/0.4 | 3.3 (-42, -24, -23) / 4.2 (19, -88, -11) |
| | Superior Frontal Gyrus | 10, 11 | 0.3/4.7 | 3.2 (-22, 46, 16) / 4.2 (25, 52, 0) |
| | Inferior Occipital Gyrus | 17, 18 | 0.0/0.4 | -999.0 (0, 0, 0) / 3.6 (25, -88, -11) |
| **WM 11** | | | | |
| | Extra-Nuclear | * | 2.0/1.9 | 4.7 (-19, -12, 17) / 5.7 (18, -12, 17) |
| | Thalamus | * | 0.8/0.9 | 5.0 (-19, -15, 16) / 5.5 (19, -15, 16) |
| | Precuneus | 7, 19 | 1.5/2.3 | 4.6 (-28, -67, 36) / 5.2 (22, -72, 45) |
| | Superior Parietal Lobule | 5, 7 | 0.3/0.8 | 3.5 (-19, -41, 59) / 5.0 (24, -69, 46) |
| | Caudate | * | 0.2/0.4 | 3.9 (-12, -8, 18) / 5.0 (16, -9, 18) |
| | Postcentral Gyrus | 2, 3, 4, 5 | 1.1/1.5 | 4.7 (-13, -42, 64) / 4.5 (16, -39, 63) |
| | Uncus | 20 | 0.4/0.1 | 4.6 (-37, -10, -29) / 3.4 (34, -6, -33) |
| | Sub-Gyral | 20 | 1.4/0.6 | 4.5 (-37, -9, -26) / 4.4 (16, -39, 60) |
| | Insula | 13 | 0.5/0.1 | 4.3 (-33, -5, 10) / 3.6 (33, -2, 11) |
| | Lentiform Nucleus | * | 1.6/1.5 | 4.3 (-30, -7, 7) / 4.2 (30, -2, 8) |
| | Lateral Ventricle | * | 0.1/0.6 | 3.6 (-7, -5, 15) / 4.2 (12, -6, 18) |

| Network 1 | Area | Brodmann Area | Volume (cc) | Peak (x, y, z) |
|---|---|---|---|---|
| | Fusiform Gyrus | 20 | 0.4/0.0 | 4.2 (-40, -10, -23) / -999.0 (0, 0, 0) |
| | Claustrum | * | 0.0/0.2 | 4.1 (-33, -2, 8) / 3.7 (30, -4, 14) |
| | Inferior Temporal Gyrus | 20, 21 | 0.3/0.0 | 3.9 (-40, -10, -29) / 3.1 (33, -5, -35) |
| | Superior Frontal Gyrus | 6, 8 | 0.2/0.4 | 3.3 (-9, 39, 44) / 3.6 (10, 22, 54) |
| | Inferior Frontal Gyrus | 45 | 0.0/0.2 | -999.0 (0, 0, 0) / 3.5 (49, 25, 11) |
| | Medial Frontal Gyrus | 6 | 0.4/0.0 | 3.4 (-12, -5, 55) / -999.0 (0, 0, 0) |
| | Cerebellar Tonsil | * | 0.2/0.0 | 3.3 (-16, -53, -40) / 3.1 (15, -54, -40) |

**Table 4. Brain regions of Network 2**

| Network 2 | Area | Brodmann Area | Volume (cc) | Peak (x, y, z) |
|---|---|---|---|---|
| **GM12** | | | | |
| | Lingual Gyrus | 17, 18 | 3.2/1.9 | (3, -88.5, -10)/(13.3, -84, -10) |
| | Sub-Gyral | * | 2.7/1.9 | (0, -85.5, -13)/(30.4, -65, -39) |
| | Inferior Occipital Gyrus | 17, 18 | 0.2/0.2 | (-10.5, -93, -15)/(0, 0, 0) |
| | Declive | * | 0.2/0.3 | (4.5, -85, -25)/(0, 0, 0) |
| | Fusiform Gyrus | 18 | 0.4/0.6 | (18.9, -21, -1)/(0, 0, 0) |
| | Inferior Parietal Lobule | 40 | 0.8/0.5 | (-43.5, -45, 45)/(0, 0, 0) |
| | Superior Parietal Lobule | 7 | 0.0/0.1 | (-27.7, 43, -48)/(0, 0, 0) |
| | Precentral Gyrus | 6 | 0.4/0.3 | (-27.1, 54, 15)/(0, 0, 0) |
| | Inferior Frontal Gyrus | 9 | 0.0/0.2 | (0, 0, 0)/(37.5, 45, 0) |
| | Precuneus | 7, 31 | 0.2/0.3 | (12.7, -75, -45)/(0, 0, 0) |
| | Insula | 13 | 0.7/0.0 | (-36, -37.5, 18)/(0, 0, 0) |
| | Middle Frontal Gyrus | 6 | 0.7/0.0 | (-27, -45, 15)/(0, 0, 0) |
| | Superior Temporal Gyrus | 39, 41 | 0.3/0.0 | (-36, -37.5, 15)/(0, 0, 0) |
| | Cuneus | 7, 17, 30 | 0.3/0.1 | (-18.7, -75, 72)/(0, 0, 0) |
| | Posterior Cingulate | 30 | 0.2/0.0 | (-96.5, -67.5, 0)/(6, 0, 0) |
| | Transverse Temporal Gyrus | 41 | 0.2/0.0 | (36.5, 12, 0)/(0, 0, 0) |

| Network 2 | Area | Brodmann Area | Volume (cc) | Peak (x, y, z) |
|---|---|---|---|---|
| **WM 5** | | | | |
| | Lingual Gyrus | 17, 18 | 4.9/0.0 | 10.3 (-10, -92, -2)/3.2 (0, -90, -3) |
| | Cuneus | 17, 18, 19 | 6.5/0.0 | 9.9 (-16, -91, 6)/3.0 (25, -93, 0) |
| | Middle Occipital Gyrus | 18, 19 | 2.1/0.0 | 9.2 (-13, -95, 16)/-999.0 (0, 0, 0) |
| | Inferior Occipital Gyrus | 17, 18, 19 | 2.1/0.0 | 7.9 (-19, -94, -7)/-999.0 (0, 0, 0) |
| | Sub-Gyral | * | 0.2/0.0 | 6.7 (-13, -101, 13)/-999.0 (0, 0, 0) |
| | Superior Frontal Gyrus | 10, 11 | 0.0/2.0 | -999.0 (0, 0, 0)/5.2 (19, 49, -13) |
| | Middle Frontal Gyrus | 9, 10, 11 | 0.1/1.5 | 3.3 (-30, 46, -10)/4.8 (24, 43, -8) |
| | Middle Temporal Gyrus | 39 | 0.6/0.0 | 4.1 (-50, -58, 7)/3.0 (42, 3, -32) |
| | Culmen | * | 0.5/0.0 | 4.0 (-39, -47, -28)/-999.0 (0, 0, 0) |
| | Precuneus | 7 | 0.8/0.0 | 3.9 (-15, -66, 48)/-999.0 (0, 0, 0) |

**Figures**

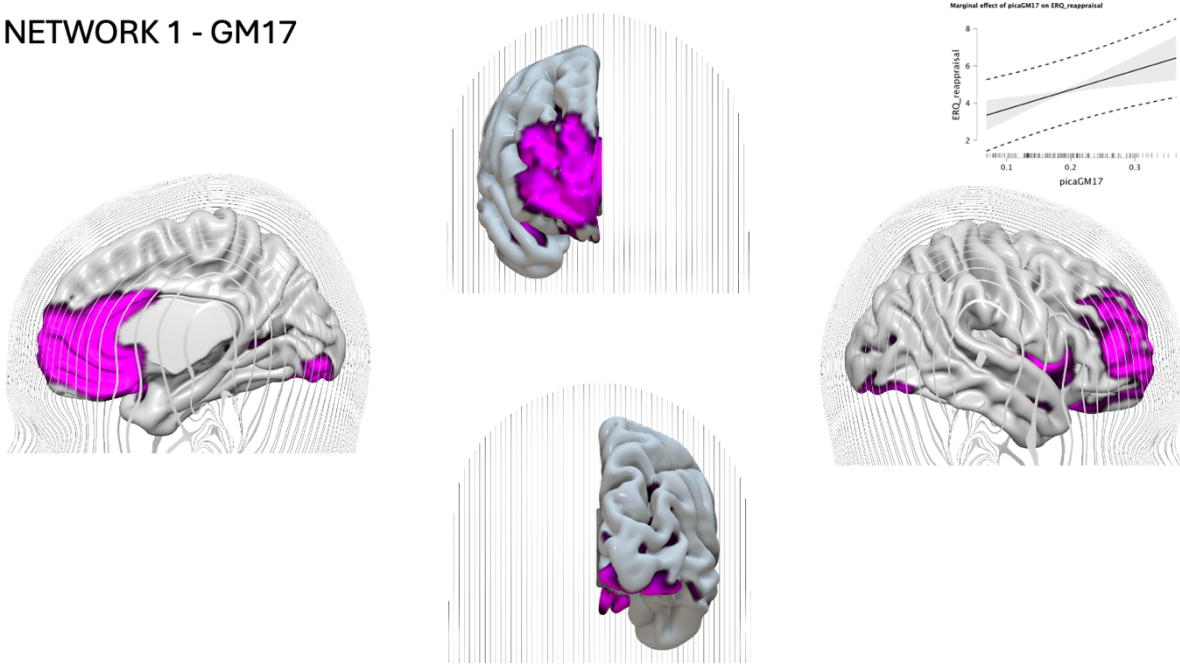

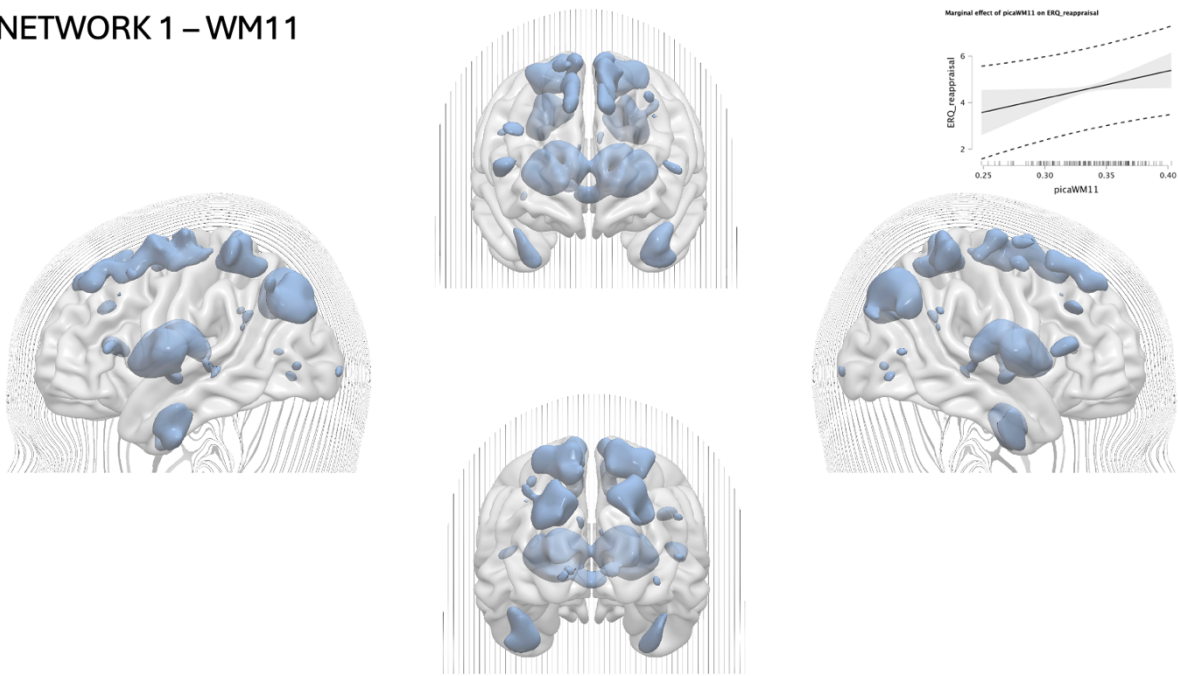

**Fig 1. Network 1 including GM17 and WM 12 predicting reappraisal.**

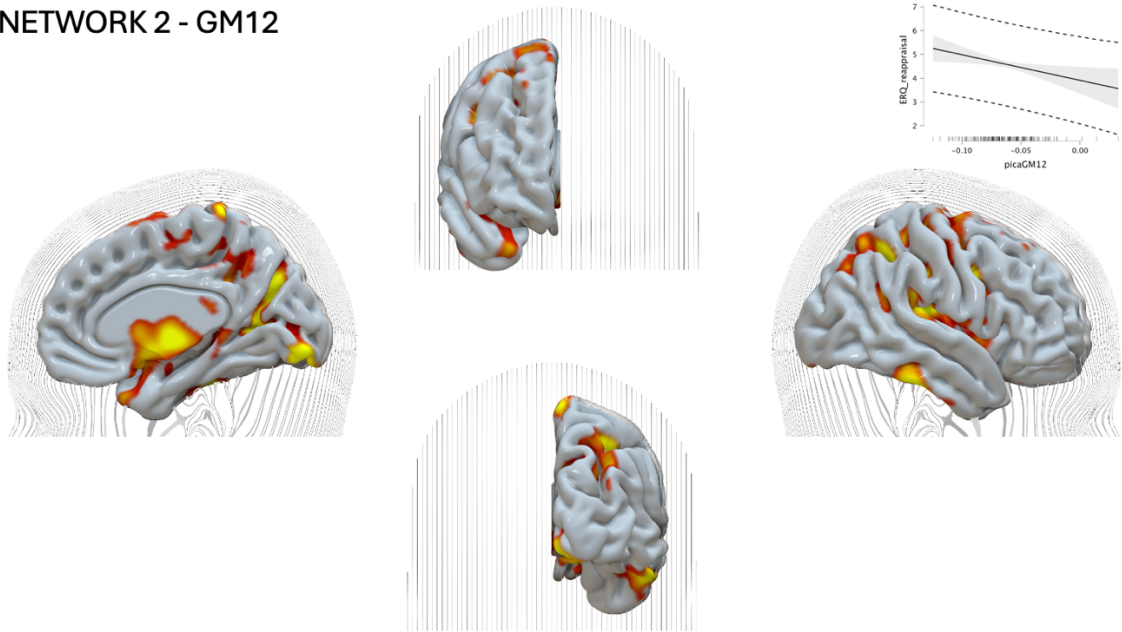

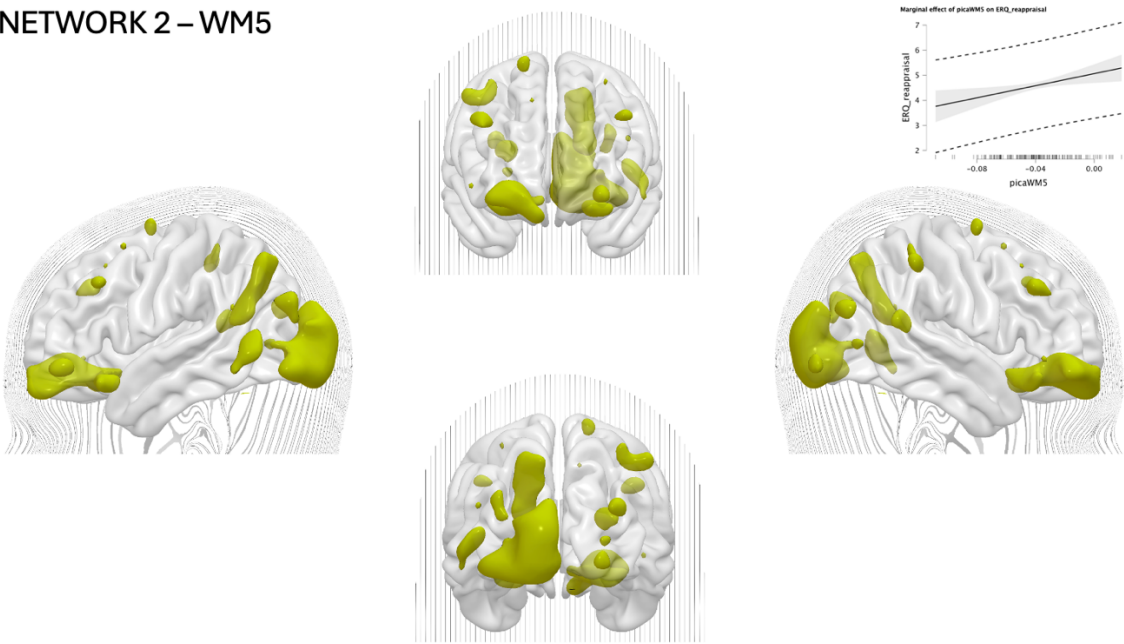

**Fig 2. Network 1 including GM12 and WM5 predicting reappraisal.**